# Does Physics Need 'Dark Matter'?

Jeremy Dunning-Davies,
Department of Physics,
University of Hull,
Hull,
England.
HU6 7RX.

email address: j.dunning-davies@hull.ac.uk

**Abstract.**

To fully understand the present position concerning so-called dark matter, it is necessary to examine the historical background since, only by following this approach, do all the pieces of the puzzle fall into place. Here an attempt is made to do this briefly and it is found that an interesting and important question is raised. This question relates to the position of electromagnetism in astronomical considerations since history indicates that, in the years following the beginning of the $20^{th}$ century, interest in electromagnetic effects appeared to wane. Hence, following an examination of the history and the presently accepted position where reliance for solutions seems confined to examining gravitational effects, attention is turned to hypotheses based on plasma physics to see if a more feasible solution to the problem of the missing mass can be furnished utilising its fundamental ideas. While the purpose here is to concentrate on dark matter and the supposed need for its introduction into physics, this consideration of electromagnetic effects combined with the realisation that most matter in the Universe is in the form of plasma also indicates alternative routes to seeking solutions for other puzzling astronomical phenomena.

## Introduction.

Some years ago, in an entirely different context, Sir Winston Churchill advised people to learn from the lessons of history. This advice should be seen as applying to all fields of human activity and now, near the beginning of the 21$^{st}$ century, it seems highly appropriate to follow this advice in several areas of physics which, at least in the eyes of some people, seem to be almost stagnating. However, here concern lies with the notion of dark matter and problems associated with both it itself and its introduction into physics. It might be remembered immediately that dark matter was introduced in order to explain some hitherto unexplainable observational results in astrophysics. The 'matter' as such is unseen and its presence is merely inferred in order to offer a seemingly reasonable explanation for some observations of a dynamical nature purely in terms of gravitational interactions. However, bearing in mind Churchill's words, first consider the background to the introduction of this mysterious matter into physics before proceeding to examine its worth and need. It is only then, possibly, that a complete picture can begin to emerge.

## Background.

As is well known, the model of the universe accepted by cosmologists today began with a suggestion made by Copernicus in the sixteenth century and developed over the next 150 years or so by Kepler, Galileo and Newton. However, the Copernican model itself developed from much earlier attempts to explain the apparent movements of the sun, moon and stars across the sky. Although Aristarchus had proposed a model of the universe with the sun at its centre as far back as the third century A.D., a model with the earth at the centre, proposed by Ptolemy, was the one generally accepted until Copernicus revived the heliocentric, or sun-centred, model. Needless to say, Copernicus's ideas met an extremely hostile reception but, over the following years, more and more accurate observations were made which showed him to be correct in his assumptions. These all started with the Danish astronomer Tycho Brahe and were continued by his well-known assistant Johann Kepler. It was Kepler who abandoned the notion that the planets move in circular orbits. Instead he showed that the orbits of the planets are described by ellipses, each with the sun at a focus. This is the basis of Kepler's first law of planetary motion. At the time he announced this result, he also revealed his second law which states that, in any given period of time, an imaginary line joining the sun and a particular planet always sweeps out an equal area. These two laws, which appeared in 1609, enabled Kepler to make accurate predictions concerning the positions of the planets. He continued with his observations, however, and ten years later announced his third law of planetary motion; that is, for any planet, the square of the period of its orbit is proportional to the cube of its distance from the sun. These three laws are still accepted but, although he was able to describe how the planets moved, Kepler was unable to explain why they moved as they did.

The next stage of the discovery process came with Galileo, who had heard how a Dutch lens grinder had used two lenses to make distant objects look larger and closer. He used this idea to study the moon, planets and stars. He was the first to observe the mountains on the moon; he noted that the Milky Way is composed of a great many faint stars which are at great distances from us; he identified four satellites of Jupiter; he noted that Venus goes through a sequence of phases; and also made important advances in our knowledge of motion. It was Galileo's ideas that were developed greatly by Isaac Newton, who was then in a position to explain why the planets move in accordance with Kepler's laws.

Building on Galileo's work, Newton was able to formulate his well-known laws of motion as applied to bodies on the earth and also to show that, for a body of mass $m$ to move in a circular path of radius $r$ with a speed $v$, a force $mv^2/r$, directed towards the centre of the circle, is required.

Newton then broke with tradition and applied these laws to the motion of the heavenly bodies. He quickly proved that any spherical body moving in accordance with Kepler's second law must be acted upon by a 'central' force; that is, in this case, a force acting along the line joining the heavenly body with the sun. He showed further that, for a body moving in an elliptical orbit with the centre of force at one focus, this force must be inversely proportional to the square of the distance between the centres of the two bodies concerned. He then made the revolutionary suggestion that, in the solar system, this centrally directed force is a **gravitational** force - just as the gravitational attraction of the earth causes an object to fall to the ground when it is dropped, so the gravitational attraction of the sun keeps a planet in its elliptical orbit. He proposed that this gravitational force between two bodies of masses $m$ and $M$, distance $d$ apart, is
$$F = GmM/d^2,$$
where $G$ is the **universal constant of gravitation**.

Newton's theories have been found, by and large, to be correct although he himself was unable to show this where planetary motion is concerned because he was ignorant of the values of the mass of the sun and of the universal constant of gravitation. It was almost a century after Newton before Henry Cavendish, using an experiment suggested by John Michell, first succeeded in measuring $G$. Once this information became available, it became possible to calculate the mass of the sun from knowledge of the distance and period of any planet, and to find the mass of any planet which has a satellite from the period and distance of that satellite. The masses of planets which do not have satellites may be found by the small gravitational effects, or perturbations, they produce on other planets. Observations of such perturbations in the orbit of Uranus led to the discovery, in 1846, of Neptune; observed further perturbations in the orbits of both Uranus and Neptune led to the discovery, in 1930, of Pluto.

Newton's theory has been eminently successful in describing and predicting motion at all levels from the microscopic to the macroscopic. However, even when theoreticians were using the theory in the mid-eighteen hundreds to predict the existence of Neptune, they were aware of the fact it didn't explain the behaviour of Mercury, the innermost of the planets, perfectly. Again, as is well known, it is generally accepted that this problem was eventually solved by utilising Einstein's General Theory of Relativity, although other legitimate solutions relying only on Newton's theory have been put forward. Whichever theory is used to explain the anomaly though, it depends on gravitational attraction alone. In fact, virtually all cosmological phenomena are explained in terms of gravitational forces and it is for this precise reason that the idea of 'dark matter' has been introduced into physics.

## More recent developments.

Most of the literature seems to indicate that one of the first issues, if not the first, to raise thoughts of the presence of 'dark matter' for providing an acceptable explanation was the problem associated with the so-called rotation curves of galaxies, those curves which plot rotation speed against distance from the galactic centre. From such curves, the mass within any given radius then follows from Newton's laws. However, as far as the Milky Way is concerned, if all the mass of the Galaxy was contained within the limits of the visible structure, Newton's laws of motion indicate that the orbital speed of both stars and gas beyond roughly 15kpc would decrease with increasing distance from the centre of the Galaxy. However, this is not what the rotation curve shows. Rather the curve, after initially rising sharply as expected, tends to level off and does not decline, seemingly implying that the amount of mass contained within successive larger radii continues to grow and this is felt to happen for distances well beyond the orbit of the Sun, possibly out to distances of as much as 50kpc. Investigation of the rotation curves of other galaxies and, indeed, of clusters of galaxies indicates that this sort of behaviour is not unusual.

The extra mass necessary to explain these results is certainly not readily visible and so the idea of so-called dark matter has been introduced to offer an explanation; an explanation, though, totally dependent on gravity and not allowing for the effects of any other possible forces which might be acting. Some point out that this is not, at first sight at least, a totally unreasonable explanation since perturbations in planetary orbits – specifically Uranus - were tentatively explained initially by postulating, by using Newtonian gravitational theory, the presence of another planet and, as a result for example, the planet Neptune was found. In a sense, until observed, Neptune might have been deemed an example of dark matter. Hence, the more modern idea of dark matter might be felt to include matter not yet observed and this could include such objects as brown dwarfs. However, it is not generally felt that such matter could account for all the apparently missing mass required to account for the problematical rotation curves.

Rather than introduce the hypothetical dark matter, Milgrom[1] investigated the possibility of modifying Newton's law of gravitation. Again, this approach cannot be immediately dismissed out of hand. After all, Newton found his law to apply to matter on the Earth. He then took the enormous intellectual step of *assuming* it to apply within the Solar System. This tremendous assumption was found to lead to consistent results; he was able to give a firm theoretical foundation for Kepler's laws and make predictions, which were later verified, concerning the Solar System. It was then natural to assume this law applied within the Universe as a whole, but such a step was yet another huge assumption. Hence, Milgrom's suggested approach was understandable. However, when all is said and done, it is a purely mathematical solution to the problem; what is the physical justification? Once again, scientists are faced with mathematics tending to rule thought, rather than remaining the tool it should be when discussing physics' problems. It is physics and purely physical reasoning which should be used to find the solution to any physics' problem. In the present context, it seems that some factors which could conceivably influence the situation are being ignored. By this is meant the fact that most matter in the Universe, unlike that with which everyone is familiar here on Earth, is in the form of plasma. Within this plasma there are likely to be magnetic fields and electric currents present and it must always be remembered that the electromagnetic force is of the order of 39 orders of magnitude greater than the gravitational force. Could this force be the source of the solution to this, and possibly other, problems? However, before considering this, a quite specific problem associated with the currently accepted idea of 'dark matter' as the saviour of astronomical observations will be examined.

## A digression on a specific 'dark matter' problem.

A worrying trend occasioned by relying on this unknown quantity termed 'dark matter' to help explain puzzling phenomena is exemplified by a recent article[2] which attempted to place direct limits on the mass of earth-bound 'dark matter'. In the article concerned, it is claimed that a method is introduced for calculating the maximum amount of dark matter that must be present in the space between the Laser Geodynamics Satellites and the Moon's orbit. The method suggested is deceptively simple. The author indicates that this quantity of dark matter is given by subtracting the values of the product of the universal constant of gravitation and the mass for the earth and the moon from the value of the same product for the two combined. This is summarised in the equation

$$GM_{dm} = GM_{combined} - GM_e - GM_m.$$

Published data is then used to give an estimate of the expected answer.

Several queries may be raised immediately, the first concerns the alternative method for determining the moon's mass by studying the orbit of a close passing asteroid which is influenced by both the gravitational field of the moon and that of the earth. It is pointed out that, from such an analysis, one may find an accurate figure for the ratio

$$R = \frac{GM_e + G\Delta M_e}{GM_m + G\Delta M_m},$$

where $\Delta M_e$ and $\Delta M_m$ denote possible contributions from earth-bound and moon-bound dark matter. Hence, to a first approximation

$$R \simeq \frac{GM_e}{GM_m}(1 + \delta)$$

where $\delta = \frac{\Delta M_e}{M_e}$. Hence,

$$GM_m \simeq \frac{GM_e}{R}(1 + \delta).$$

Due to this relationship, one must enquire as to the range of validity of some subsequent equations. Also, it has to be noted that any claims following must be of dubious validity since, to arrive at them, so many approximations have been made. Therefore, the claim of a 'potential one percent accuracy' must be open to doubt.

A numerical evaluation then follows and this raises even more serious queries. Firstly, each of the separate values of the product $GM$ is found by different methods, each involving different sets of assumptions. The figures are then manipulated in different ways, again with implicit assumptions, before the final calculation. After all this, the $GM$ for dark matter is found to be 0.0001±0.0016 which, on dividing by the value of $GM$ for the earth leads to a ratio of (0.3±4)×10$^{-9}$. Based on this, it is asserted that there must be a mass of dark matter less than 4×10$^{-9}$ times the mass of the earth in the volume of space considered – assuming $G$ constant.

There are at least two problems with the conclusion above. The first concerns the statistical significance of 0.0001±0.0016. The second concerns the assertion that there must be a mass of dark matter less than 4×10$^{-9}$ times the mass of the earth in the volume of space considered. Consider each in turn.

The value used for the combined Earth-Moon GM is 403,503.2357 ± 0.0014. If for comparison the separate Earth and Moon values are added and standard interval calculations used to obtain the new error, 403,503.2356 ± 0.0011 results. If these two values with error bars are viewed on a chart, the second summed value fits perfectly within the 95% confidence error bars of the original combined value. There is no significant difference between the two values. Yes, some misguided mathematical calculations may be performed to derive a difference value of 0.0001 ("in the noise" so to speak), but the figure is not meaningful.

Consider now the ratio $(0.3\pm4)\times10^{-9}$. This value is used to assert there is at most $4\times10^{-9}$ times the mass of the earth in the volume of space considered. However, it follows that -3.7 is as statistically valid as +4.3. The best that may be deduced is that there is a 95% likelihood of there being, or not being, any dark matter in the stated volume of space.

Whatever one's belief on the existence, or not, of dark matter, probably the most important comment in the paper occurs in a footnote, where the author comments that the analysis in the paper is based on purely gravitational considerations. It must always be remembered that other forces, such as the electromagnetic force, could be exerting influences also. Also, it is worrying that figures such as those discussed here are not simply dismissed as being insignificant by scientists; in fact, they seem to be given a degree of credence.

**Electromagnetic ideas.**

Again noting Churchill's advice, it is interesting to note that, following the introduction of Newton's mechanical ideas, work still proceeded apace investigating electromagnetic phenomena and this continued at least into the earlier years of the twentieth century, as is evidenced by the contents of J. J. Thomson's book *Electricity and Matter*[3], which is concerned with a series of lectures he gave at Yale University in May, 1903. However, this book provides but one example to illustrate the very real emphasis on work involving the effects of the electric and magnetic fields, work which constantly sought an explanation in terms of those forces for the concept of mass. As an aside, it is possibly worth noting that, in this book, Thomson talks routinely of the equivalence of mass and energy, and that in lectures delivered in 1903. However, after those early years of the century, the emphasis seems to have shifted to explanations of phenomena purely in terms of gravitational effects as far as most mainline research has been concerned. Considering that it is accepted that much of the matter in the Universe is in the form of plasma, this seems a retrograde step and this view is surely strengthened when the work of such as Kristian Birkeland and Hannes Alfvén is concerned. One may only speculate as to why the emphasis of much scientific research changed in this way. However, thanks to people like Birkeland, Alfvén and (more recently) Peratt, work in the areas of electromagnetism and plasma physics did continue and it should be noted that much of the work on plasmas has been via laboratory experiments, so hard experimental evidence is available to support any claims made.

The work on plasmas and other electromagnetic phenomena has inspired some people to examine astronomical phenomena in these terms and this has resulted in the so-called Electric Universe idea as expounded, for example, in the books *The Electric Universe* by Wallace Thornhill and David Talbott[4] and *The Electric Sky* by Donald Scott[5]. Reading through this material makes one immediately aware that, while like orthodox accepted theory, the electric universe ideas are supported by much computer modelling, it can also draw on parallels in astronomy with plasma phenomena observed in the laboratory. Admittedly, drawing such parallels involves scaling up tremendously but assuming this possible is little different from assuming that laws seemingly applicable here on the Earth are also applicable in the Solar System and, indeed, throughout the Universe. However, at least visually, some of the phenomena observed in the laboratory are very like what is observed by some of the most powerful of telescopes; - electric currents in plasma naturally form filaments due to the so-called 'pinch effect' of the induced magnetic field. Electromagnetic interactions cause these filaments to rotate about one another to form a helical 'Birkeland Current' filament pair and this is very much the structure seen in the Double Helix nebula near the galactic centre; again, the Hubble image of the planetary nebula NGC6751 looks remarkably like the view down the barrel of a plasma focus device. Examples such as these prove nothing but should awaken people to the possibility of alternative explanations for astronomical phenomena.

Much of the laboratory work seems to have originated with the work of Kristian Birkeland more than one hundred years ago. It was during his Arctic expeditions at the end of the 19$^{th}$ century that the first magnetic field measurements were made of the Earth's polar regions. His findings also indicated the likelihood that the auroras were produced by charged particles originating in the Sun and guided by the Earth's magnetic field. Birkeland, though, was an experimentalist and is still known for his Terrella experiments carried out in a near vacuum and in which he used a magnetised metallic sphere to represent the Sun or a planet and subjected it to electrical discharges. By this means, he was able to produce scaled down auroral-type displays as well as analogues of other astronomical phenomena. These claims, however, were only vindicated finally by satellite measurements in the 1960's and 70's. To that point in time, his experimental and observational achievements had tended to be overshadowed by the purely theoretical predictions and explanations of the geophysicist, Sydney Chapman. Once again, powerful mathematics seems to have held sway over the more expected techniques of physics – experimentation and observation, with mathematics a mere tool to be used when necessary. This is not to decry Chapman's work but to emphasise the overwhelming importance of the physics when investigating natural phenomena.

Birkeland also showed experimentally that electric currents tend to flow along filaments shaped by current induced magnetic fields. Of course, this confirmed observations of Ampère that indicated that two parallel currents flowing in wires experience a long range attractive magnetic force that brings them closer together. However, as plasma currents come closer together, they are free to rotate about each other. Such action generates a short range repulsive magnetic force which keeps the filaments separated so that they are, in effect, insulated from each other and able to maintain their separate identities. The end effect is for them to appear like a twisted rope and it is this configuration which is termed a 'Birkeland current', as was mentioned earlier when the Double Helix nebula was noted as a possible example. Satellites orbiting above the auroras in the 60's and 70's were able to detect a movement of ions, indicating that electric currents were present. Later missions found quasi-steady electric fields above the auroras following the magnetic field lines, thus lending some credence to Birkeland's claim of the existence of an electric circuit between the earth and the Sun. Some may be sceptical of this latter interpretation but it is undoubtedly true that much of the material in the Universe is in the form of plasma and there is certainly electric and magnetic activity occurring in abundance. This means there are numerous very good reasons for considering the effects of the electromagnetic force in the Universe, only one of which could be the resolution of the problem of the missing mass.

However, precisely what is the Electric Universe? In truth, it is really simply an hypothesis, a new way of interpreting known data by utilising both new and well-established knowledge relating to electricity and plasma. It should be emphasised immediately that, in this new interpretation, gravity still has a role to play but it is a secondary one since the electric force is so much more powerful. A major point to be stressed from the outset is that, in this interpretation of astronomical phenomena, scientists are able to call on evidence from laboratory based experiments to help form and support suggested explanations for a wide variety of phenomena. It has been found that, as explained in more detail in the above-mentioned books, an electrified plasma in a laboratory is a good model for providing possible explanations for many recently observed astronomical phenomena which, in several cases, have caused puzzlement for astronomers seeking explanations via more orthodox gravitationally based theories. This is not to say that gravity is ignored and regarded as irrelevant; rather, the possible effects of the electromagnetic force on astronomical phenomena are investigated while still recognising the importance of gravitational effects. In the electric universe, the gravitational systems of galaxies, stars, moons and planets are felt to have their origins in the proven ability of electricity to generate both structure and rotation in plasma. It is felt further that the force of gravity assumes importance only as the electromagnetic forces approach equilibrium. As has been noted already, great consternation has been caused in astronomical circles by the realisation that gravity, as presently understood, cannot explain much that is observed if the amount of mass available is as now felt to be present. Hence, instead of positing the existence of 'dark matter' or following the path of modifying Newton's well-tried law of gravitation, it is suggested here that the possible effects of the electromagnetic force be examined to see if, in conjunction with orthodox ideas on gravity, these puzzling observations can be explained.

A point which is often relegated to the background when discussing the solution of problems through the introduction of dark matter is the fact that the missing mass, if there really is any missing mass, is not absent homogeneously throughout the Universe; it is missing only in specific places - for example, in the outer regions of galaxies. Hence, possible solutions, such as the idea that neutrinos possess mass, which are essentially homogeneous in nature cannot be acceptable. It should be mentioned at this point though that, in the Electric Universe model, neutrinos do possess mass and are extremely important. They respond only weakly to massive objects such as stars and galaxies but form an extended atmosphere which, for example, refracts light around the Sun from distant stars and this offers an alternative explanation for the so-called gravitational bending of light. On the other hand, in this model, neutrinos are not required to explain galactic rotation although they must contribute to the masses of both stars and galaxies. Again, having some mass, neutrinos will not be distributed homogeneously.

However, returning to the realisation that much of the matter permeating the Universe is in the form of plasma, it might be remembered that these clouds of plasma respond to the well-known laws of Maxwell. Also, as pointed out by Scott in his book[5], another law, formulated by Lorentz, does help explain the galactic speeds alluded to earlier. This law states that
> *a moving charged particle's momentum (speed or direction) can be changed by application of either an electric field or a magnetic field or both.*

This seems a highly likely contributory factor, at least, causing galaxies to rotate as they are perceived to do but would indicate, contrary to the accepted view, that gravity has less to do with things than has been thought. However, it should be emphasised that nowhere is it being suggested that Newton's law of gravitation is wrong or in need of modification; it is simply being suggested that, in deep space where everything swims in a sea of plasma, the Maxwell – Lorentz electromagnetic forces dominate over those of gravity.

It might be remembered also that the Lorentz force alluded to here changes a charged particle's momentum and that change is directly proportional to the strength of the magnetic field through which the particle is moving. Further, the strength of a magnetic field produced by an electric current is inversely proportional to the distance from the current but the gravitational force between stars is inversely proportional to the *square* of the distance. This well-known difference between the two forces could lie at the heart of the problem of the galactic rotation curves; certainly it seems an avenue worth exploring further, especially considering the fact that more and more space missions are indicating that electromagnetic forces are distributed more widely throughout space and are, of course, many orders of magnitude stronger than gravitational forces.

Much time, effort and money is spent worldwide on producing elaborate computer programs which purport to support the prevailing belief in the Big Bang as being the correct theory explaining how the Universe originated. However, as well as a great many laboratory experiments being performed to help establish plasma properties[6], it has been shown also, using the Maxwell and Lorentz equations, that streams of charged particles, such as are found in the intergalactic plasma, will evolve into the familiar galactic shapes under the influence of electromagnetic forces. The results fit extremely well with the observed velocity profiles in the galaxies and all this with no recourse to missing mass. Much of this simulation work has been carried out by Anthony Peratt and is reported in various issues of the IEEE Transactions on Plasma Science, a highly prestigious journal.

## Conclusion.

Dark matter and other associated topics such as dark energy, the missing neutrinos, the place of string theory in physics, amongst others, are huge topics, each occupying a vast place in the scientific literature. Here a brief overview of the situation in one, dark matter, is presented for reflection and contemplation. The historical approach has been adopted because of the nature and development of the subject. It does appear that, during the last one hundred years, emphasis has shifted almost completely to gravity when seeking explanations for observed phenomena. This may not be so but it certainly appears to be the case and the possible effects of the much more powerful electromagnetic force seem relegated to the background of scientific investigations, particularly in the fields of astronomy/astrophysics and cosmology. However, as noted, more and more information is being collected by satellites indicating that, as might be expected given that so much of the matter in the Universe is in the form of plasma, there is a great deal of electric and magnetic activity taking place in all the space pervading our Universe. Here, by looking at just one problem, it is seen that investigating the effects of this enormously strong force, the electromagnetic force, in our Universe could conceivably help solve several present day scientific mysteries and assist in extending the boundaries of human knowledge in yet another huge leap forward.

In his Nobel Lecture of December 1970, Hannes Alfvén said, "The centre of gravity of the physical sciences is always moving. Every new discovery displaces the interest and the emphasis." Maybe those working in the fields of astronomy/astrophysics and cosmology especially should take note of these words of wisdom uttered by an acknowledged scientific thinker and open their minds to the possibilities highlighted here of the importance of the electromagnetic field in helping solve problems in their preferred scientific domains.

## Acknowledgement.

The author wishes to thank Dr. Thomas Wilson for invaluable collaboration when originally discussing the material contained in the article purporting to place direct limits on the mass of earth-bound 'dark matter'. Also, thanks are due to Dr. Wallace Thornhill for reminding the author that, in the Electric Universe model, neutrinos do possess mass.

**References.**